\documentclass[prl,showpacs,twocolumn,aps]{revtex4}

\usepackage{amsmath}
\usepackage{amssymb}
\usepackage{latexsym}
\usepackage{amsfonts}
\usepackage{epsfig}
\usepackage{psfrag}
\usepackage{graphicx}

\newcommand{\ket}[1]{\left|#1\right>}
\newcommand{\bra}[1]{\left<#1\right|}   

\newcommand{\nn}{\nonumber\\}

\newcommand{\f}[1]{\mbox{\boldmath$#1$}}
\newcommand{\fk}[1]{\mbox{\boldmath$\scriptstyle#1$}}

\newcommand{\bea}{\begin{eqnarray}}
\newcommand{\ea}{\end{eqnarray}}
\newcommand{\eea}{\end{eqnarray}}
\newcommand{\ord}{\,{\cal O}}
\newcommand{\li}{\,\widehat{\cal L}}
\newcommand{\tr}{\,{\rm Tr}}

\begin{document}

\title{Emergence of coherence in the Mott--superfluid quench of 
the Bose-Hubbard model} 

\author{Patrick Navez and Ralf Sch\"utzhold$^*$}

\affiliation{
Fakult\"at f\"ur Physik, Universit\"at Duisburg-Essen, 
Lotharstrasse 1, 47057 Duisburg, Germany}

\date{\today}

\begin{abstract}
We study the quench from the Mott to the superfluid phase in the 
Bose-Hubbard model and investigate the spatial-temporal growth of phase 
coherence, i.e., phase locking between initially uncorrelated sites.
To this end, we establish a hierarchy of correlations via a controlled 
expansion into inverse powers of the coordination number $1/Z$. 
It turns out that the off-diagonal long-range order spreads with a 
constant propagation speed, forming local condensate patches, whereas 
the phase correlator follows a diffusion-like growth rate.  
\end{abstract}

\pacs{
03.75.Gg, 
03.75.Kk, 
05.30.Rt, 
03.75.Lm. 
}

\maketitle

A sweep through a symmetry-breaking quantum phase transition is one 
of the simplest ways to create and amplify quantum correlations, 
i.e., entanglement.  
As the initial quantum state is symmetric, all directions of 
symmetry breaking are equally likely and seeded by quantum 
fluctuations. 
Furthermore, the diverging response time at the critical point 
indicates that the many-particle quantum system is driven far 
away from equilibrium during the sweep. 
While nearby points will most likely break the initial symmetry 
in the same direction, two very distant points may spontaneously 
select different directions of symmetry breaking \cite{Kibble-Zurek}. 
As a result, the spatial order parameter distribution after the 
quench will be inhomogeneous -- and its spatial correlations are 
directly determined by quantum correlations 
\cite{entanglement}. 

In contrast to the static properties of classical and quantum phase 
transitions, much less is known about their non-equilibrium dynamics.  
Motivated by recent experiments \cite{bec,lattice}, as well as from 
a more fundamental point of view, this field of research is attracting 
increasing interest \cite{interest}. 
Open questions in this context include: 
How is the new order parameter established and how fast does it spread?  
Are there universal scaling laws \cite{universal} similar to those in 
static phase transitions close to the critical point? 
In this Letter, we address these questions for the Bose-Hubbard model
which is considered \cite{Sachdev} one of the prototypical examples for 
quantum phase transitions and also relevant for experiments 
in optical lattices \cite{lattice}.
%
%
The Bose-Hubbard Hamiltonian is given by ($\hbar=1$)
\bea
\label{Hamiltonian}
\hat H
=
-\frac{J}{Z}\sum_{\mu,\nu} T_{\mu\nu} \hat a^\dagger_\mu \hat a_\nu
+\frac{U}{2}\sum_{\mu} \hat n_\mu(\hat n_\mu-1)
\,,
\ea
where $\hat a^\dagger_\mu$ and $\hat a_\nu$ are the 
creation and annihilation operators at the lattice sites 
$\mu$ and $\nu$, respectively. 
The lattice structure is encoded in the tunneling matrix
$T_{\mu\nu}\in\{0,1\}$ and $J$ denotes the hopping rate. 
The coordination number $Z=\sum_{\nu} T_{\mu\nu}\in\mathbb N$ 
counts the number of tunnelling neighbors at any given site $\mu$.
Finally, $U$ is the on-site interaction with 
$\hat n_\mu=\hat a^\dagger_\mu \hat a_\mu$. 
For simplicity, we assume an average filling of one boson 
per site $\langle\hat n_\mu\rangle=1$.

At a critical ratio of $J/U$ (see below), this model (\ref{Hamiltonian}) 
features a quantum phase transition separating the symmetric Mott 
insulator phase with a gap from the superfluid phase with long-range 
order (i.e., broken symmetry) and Goldstone modes \cite{Fisher}. 
We start deep in the Mott phase $J=0$ with 
$\ket{\Psi_{\rm Mott}}=\prod_\mu\hat a^\dagger_\mu\ket{0}$ 
and quench to the superfluid regime by means of an instantaneous 
switching to a finite value of $J$. 
The theoretical description of the ensuing non-equilibrium many-particle 
dynamics associated with this process is a non-trivial task for the 
non-integrable Bose-Hubbard Hamiltonian (\ref{Hamiltonian}).
There are two major options: numerical computations which are limited to 
systems of finite size or analytical calculations which require suitable 
approximations. 
In order to control the accuracy and consistency of these approximations, 
they should be based on the expansion in terms of a small or large 
parameter. 
For the Bose-Hubbard model, this could be a large filling
$\langle\hat n_\mu\rangle\gg1$ \cite{large-N} or a large number 
of interaction partners for the extended version \cite{supersolid}. 
In this Letter, we assume a large coordination number $Z\gg1$ 
and employ a systematic expansion in $1/Z$.
A large $Z$ can occur in a large number of spatial dimensions or 
a large number of tunnelling partners.
In the following, we focus on the second case and assume two spatial 
dimensions for simplicity.  

Let us consider the reduced density matrices for one lattice site 
$\hat\rho_\mu=\tr_{\not\mu}\{\hat\rho\}$ 
and for two sites 
$\hat\rho_{\mu\nu}=\tr_{\not\mu\not\nu}\{\hat\rho\}$ etc.
Furthermore, we separate the correlated parts via  
$\hat\rho_{\mu\nu}=\hat\rho_{\mu\nu}^{\rm c}+\hat\rho_{\mu}\hat\rho_{\nu}$
as well as 
$\hat\rho_{\mu\nu\lambda}=\hat\rho_{\mu\nu\lambda}^{\rm c}+
\hat\rho_{\mu\nu}^{\rm c}\hat\rho_{\lambda}+
\hat\rho_{\mu\lambda}^{\rm c}\hat\rho_{\nu}+
\hat\rho_{\nu\lambda}^{\rm c}\hat\rho_{\mu}+
\hat\rho_{\mu}\hat\rho_{\nu}\hat\rho_{\lambda}$
etc.
Our derivation is based on the following scaling hierarchy of 
correlations 
\bea
\label{hierarchy}
\hat\rho_{\cal S}^{\rm c}=\ord\left(Z^{1-|\cal S|}\right)
\ea
where $|\cal S|$ is the number of lattice sites in the set $\cal S$, 
i.e., 
$\hat\rho_\mu=\ord(Z^0)$, 
$\hat\rho_{\mu\nu}^{\rm c}=\ord(Z^{-1})$, 
$\hat\rho_{\mu\nu\lambda}^{\rm c}=\ord(Z^{-2})$
etc.
This hierarchy (\ref{hierarchy}) is similar to the quantum de~Finetti 
theorem \cite{Finetti}, the generalized cumulant expansion \cite{kubo}, 
and the BBGKY hierarchy \cite{Balescu}, but we are considering lattice 
sites instead of particles. 
In order to derive the hierarchy (\ref{hierarchy}), we introduce 
the generating functional
\bea
{\cal F}(\hat\alpha)
={\cal F}(\{ \hat\alpha_\mu \}) 
=\ln \tr \left\{ \hat\rho \prod_\mu (\mathbf{1}_\mu + \hat\alpha_\mu)\right\}
\,,
\eea 
where $\hat\alpha_\mu$ is an arbitrary operator acting on the 
lattice site $\mu$.
This functional generates all correlated density matrices via 
$\hat\alpha_\mu$-derivatives 
$\hat\rho_{\mu}=\partial{\cal F}/\partial\hat\alpha_\mu|_{\hat\alpha=0}$, 
as well as 
$\hat\rho_{\mu\nu}^{\rm c}=\partial^2{\cal F}/
\partial\hat\alpha_\mu\partial\hat\alpha_\nu|_{\hat\alpha=0}$ 
etc., 
%
%
%
where we have used the notation 
$\bra{n_\mu}\partial{\cal F}/\partial\hat\alpha_\mu\ket{m_\mu}=
\partial{\cal F}/\partial\bra{n_\mu}\hat\alpha_\mu\ket{m_\mu}$. 
Introducing the Liouville super-operators $\li_\mu$ and $\li_{\mu \nu}$
via 
$i \partial_t \hat\rho = [\hat H,\hat\rho] = 
\sum_\mu\li_\mu\hat\rho+\sum_{\mu,\nu} \li_{\mu \nu}\hat\rho/Z$, 
%
%
%
%
the temporal evolution of ${\cal F}$ is given by 
\begin{widetext}
\bea
\label{eqgf}
i \partial_t 
{\cal F}(\hat\alpha)
=
\sum_\mu\tr_\mu 
\left\{
\hat\alpha_\mu \li_\mu \frac{\partial{\cal F}}{\partial\hat\alpha_\mu}
\right\}
+
\frac{1}{Z} \sum_{\mu,\nu} \tr_{\mu\nu} 
\left\{
(\hat\alpha_\mu + \hat\alpha_\nu +\hat\alpha_\mu \hat\alpha_\nu) 
\li_{\mu \nu} 
\left(
\frac{\partial^2{\cal F}}{\partial\hat\alpha_\mu\partial\hat\alpha_\nu}
+
\frac{\partial{\cal F}}{\partial\hat\alpha_\mu}
\frac{\partial{\cal F}}{\partial\hat\alpha_\nu}
\right)
\right\}
\,.
\eea
By taking successive derivatives, we establish the following set of 
equations for the correlated density matrices 
\bea
\label{general}
i \partial_t\hat\rho^c_{\cal S}
=
\sum_{\mu \in {\cal S}} \li_\mu \hat\rho^c_{\cal S}
+
\frac{1}{Z}\sum_{\mu,\nu\in{\cal S}}
\li_{\mu \nu}\,\hat\rho^c_{\cal S} 
+ 
\frac{1}{Z}
\sum_{\kappa\notin{\cal S}} \sum_{\mu\in{\cal S}} 
\tr_{\kappa}
\left\{
\li^S_{\mu \kappa}
\hat\rho^c_{{\cal S}\cup {\kappa}}
+ 
\sum_{{\cal P}\subseteq{\cal S}\setminus\{\mu\}}
^{{\cal P}\cup\bar{\cal P}={\cal S}\setminus\{\mu\}}
\li^S_{\mu \kappa}
\hat\rho^c_{\{\mu\}\cup{\cal P}}\,
\hat\rho^c_{\{\kappa\}\cup\bar{\cal P}}
\right\}
+
\nn
\frac{1}{Z}
\sum_{\mu,\nu\in{\cal S}} 
\sum_{{\cal P}\subseteq{\cal S}\setminus\{\mu,\nu\}}
^{{\cal P}\cup\bar{\cal P}={\cal S}\setminus\{\mu,\nu\}}
\left[
\li_{\mu \nu}\,
\hat\rho^c_{\{\mu\}\cup{\cal P}}\,
\hat\rho^c_{\{\nu\}\cup\bar{\cal P}}
- 
\tr_{\nu}\left\{
\li^S_{\mu \nu}\left(\hat\rho^c_{\{\mu,\nu\}\cup\bar{\cal P}} 
+
\sum_{{\cal Q}\subseteq\bar{\cal P}}
^{{\cal Q}\cup\bar{\cal Q}=\bar{\cal P}}
\hat\rho^c_{\{\mu\}\cup{\cal Q}}\,
\hat\rho^c_{\{\nu\}\cup\bar{\cal Q}}
\right)
\right\}
\hat\rho^c_{\{\nu\}\cup{\cal P}}
\right]
\,,
\eea
where $\li_{\mu \nu}^S=\li_{\mu \nu}+\li_{\nu \mu}$.
A careful inspection of this set of equations shows that the hierarchy 
in (\ref{hierarchy}) is preserved in time. 
If all the correlated density matrixes $\hat\rho^c_{\cal P}$ 
on the r.h.s.\ obey the hierarchy 
$\hat\rho^c_{\cal P}=\ord(Z^{1-|\cal P|})$, then the time derivative 
on the l.h.s.\ does also satisfy  (\ref{hierarchy}).
Therefore, starting deep in the Mott phase 
$\ket{\Psi_{\rm Mott}}=\prod_\mu\hat a^\dagger_\mu\ket{0}$ 
where (\ref{hierarchy}) is trivially satisfied since all correlations 
vanish, we find that (\ref{hierarchy}) remains valid for a finite time; 
more precisely, for a time scale of $\ord(\ln Z)$ limited by the 
instability of the growing modes, see below.
Let us consider some examples for the above evolution equation: 
for one lattice site ${\cal S}=\{\mu\}$, we get
\bea
\label{one-site}
i\partial_t\hat\rho_{\mu}
=
\li_\mu  \hat\rho_{\mu}
+
\frac{1}{Z}
\sum_{\kappa}\tr_{\kappa}\left\{
\li^S_{\mu \kappa}
(\hat\rho^c_{\mu \kappa}+\hat\rho_\mu \hat\rho_\kappa)\right\}
\,.
\eea
Since $\hat\rho^c_{\mu \kappa}$ is suppressed by $\ord(1/Z)$, we may 
neglect this term to the lowest order in $1/Z$ and thereby obtain the 
Gutzwiller approach \cite{Navez,Gutzwiller}. 
Starting in the Mott phase $\langle\hat a_\kappa\rangle=0$, we have 
$\tr_{\kappa}\{\li^S_{\mu\kappa}\hat\rho_\kappa\}=0$ and thus we get 
$\hat\rho_{\mu}=\ket{1}_\mu\!\bra{1}+\ord(1/Z)=\hat\rho_{\mu}^0+\ord(1/Z)$. 
The two-point correlation can be studied with ${\cal S}=\{\mu,\nu\}$
\bea
\label{two-sites}
i \partial_t \hat\rho^c_{\mu \nu}
&=&
\li_\mu\hat\rho^c_{\mu\nu}
+
\frac1Z\li_{\mu\nu}
(\hat\rho^c_{\mu\nu}+\hat\rho_\mu\hat\rho_\nu)
+
\frac1Z
\sum_{\kappa\not=\mu,\nu} 
\tr_{\kappa}
\left\{
\li^S_{\mu \kappa}
(\hat\rho^c_{\mu\nu\kappa}+
\hat\rho^c_{\mu\nu}\hat\rho_{\kappa}+\hat\rho^c_{\nu\kappa}\hat\rho_{\mu})
\right\}
-
\frac{\hat\rho_{\mu}}{Z}
\tr_{\mu}
\left\{\li^S_{\mu\nu}
(\hat\rho^c_{\mu\nu}+\hat\rho_\mu\hat\rho_\nu)
\right\}
\nn
&&
+(\mu\leftrightarrow\nu)
\,.
\eea
Again, to the lowest order in $1/Z$, we may neglect the three-point 
correlation $\hat\rho^c_{\mu\nu\kappa}=\ord(1/Z^2)$ and replace 
$\hat\rho_{\mu}$ by $\hat\rho_{\mu}^0$ arriving at a closed set 
of equations for $\hat\rho^c_{\mu \nu}$. 
Furthermore, using $\tr_{\nu}\{\li^S_{\mu\nu}\hat\rho_\nu\}=0$, 
this simplifies to  
\bea
i \partial_t \hat\rho^c_{\mu \nu}
=
\left(\li_\mu+\li_\nu\right)\hat\rho^c_{\mu\nu}
+
\frac1Z\li_{\mu\nu}^S\hat\rho_\mu^0\hat\rho_\nu^0
+
\frac1Z
\sum_{\kappa\not=\mu,\nu} 
\tr_{\kappa}
\left\{
\li^S_{\mu\kappa}\hat\rho^c_{\nu\kappa}\hat\rho_{\mu}^0
+\li^S_{\nu\kappa}\hat\rho^c_{\mu\kappa}\hat\rho_{\nu}^0
\right\}
+\ord(1/Z^2)
\,.
\eea
\end{widetext}
Introducing 
$\hat p_\mu=\ket{1}_\mu\!\bra{2}$ 
and 
$\hat h_\mu=\ket{0}_\mu\!\bra{1}$ 
as local particle and hole operators,  we find that 
their correlation functions 
$f^{11}_{\mu\nu}=\tr\{\hat\rho\,\hat h_\mu^\dagger\hat h_\nu\}$, 
$f^{12}_{\mu\nu}=\tr\{\hat\rho\,\hat h_\mu^\dagger\hat p_\nu\}$, 
$f^{21}_{\mu\nu}=\tr\{\hat\rho\,\hat p_\mu^\dagger\hat h_\nu\}$, 
and 
$f^{22}_{\mu\nu}=\tr\{\hat\rho\,\hat p_\mu^\dagger\hat p_\nu\}$, 
obey a closed linear system of equations. 
Correlation functions $f^{nm}_{\mu\nu}$ containing higher occupation 
numbers $n,m\geq3$ decouple and obey a homogeneous set of equations 
without the source terms stemming from 
$\li_{\mu\nu}^S\hat\rho_\mu^0\hat\rho_\nu^0$. 
Thus, they are trivially zero assuming the Mott state initially
-- up to the accuracy $\ord(1/Z^2)$ under consideration.
%
%
%

Assuming discrete translational invariance, we may simplify the 
equations via a Fourier transformation 
\bea 
\label{f2}
(i\partial_t-U+3JT_{\fk{k}}) f_{\fk{k}}^{12}
&=&
-\sqrt{2}JT_{\fk{k}}(f_{\fk{k}}^{11}+f_{\fk{k}}^{22}+1)
\,,
\nn
i\partial_t f_{\fk{k}}^{11}=i\partial_t f_{\fk{k}}^{22}
&=&
\sqrt{2}JT_{\fk{k}}(f_{\fk{k}}^{12}-f_{\fk{k}}^{21}) 
\,,
\eea 
with $f_{\fk{k}}^{21}=(f_{\fk{k}}^{12})^*$. 
We find that $f_{\fk{k}}^{11}= f_{\fk{k}}^{22}$ indicating 
an effective particle-hole symmetry.
The evolution is fully determined by $J$, $U$ and the Fourier 
transform of the tunnelling matrix 
\bea
\label{Fourier}
T_{\fk{k}}
=
\frac1Z\sum_{\mu}T_{\mu\nu}
e^{i\fk{k}\cdot\Delta\fk{r}_{\mu\nu}}
=
1-\frac{\f{k}^2}{2m^*}+\ord(\f{k}^4)
\,.
\eea 
Assuming discrete rotational symmetry of the lattice, we obtain 
exact isotropy at small $\f{k}$ with a unique effective mass $m^*$; 
otherwise one would have $m_x^*\neq m_y^*$ etc. 
For large $Z$, the sum over lattice sites $\mu$ in (\ref{Fourier}) 
involves many terms and thus only small wavenumbers yield 
significant contributions, corresponding to the effective mass 
being small $m^*=\ord(1/Z)$.  

The linear system in (\ref{f2}) can be rewritten in matrix form 
$i\partial_t{\mathfrak f}_{\fk{k}}=
[{\mathfrak L}_{\fk{k}},{\mathfrak f}_{\fk{k}}]+ 
{\mathfrak s}_{\fk{k}}$
with ${\mathfrak f}_{\fk{k}}=(f_{\fk{k}}^{nm})$,
the source term 
${\mathfrak s}_{\fk{k}}$ 
and the Liouvillian matrix 
${\mathfrak L}_{\fk{k}}$,
whose eigenvalues $\omega^{\rm p/h}_{\fk{k}}$ 
are associated with quasi-particle/quasi-hole excitations 
\cite{Navez}.
As one would expect from the particle-hole symmetry, two of the 
four eigenfrequencies of the full linear system in (\ref{f2}) 
vanish and the other two read 
\bea
\label{omega}
\pm
\omega_{\fk{k}}
=
\omega^{\rm p}_{\fk{k}}-\omega^{\rm h}_{\fk{k}}
=
\sqrt{U^2-6JUT_{\fk{k}}+J^2T^2_{\fk{k}}}
\,.
\eea 
After the quench from $J_{\rm in}\ll U$ to $J_{\rm out}=J$,  
the set of equations in (\ref{f2}) is solved for the initial 
conditions 
$\langle\hat h_\mu^\dagger\hat h_\nu\rangle_0=\delta_{\mu\nu}$ 
and 
$\langle\hat h_\mu^\dagger\hat p_\nu\rangle_0=
\langle\hat p_\mu^\dagger\hat h_\nu\rangle_0=
\langle\hat p_\mu^\dagger\hat p_\nu\rangle_0=0$.  
Inserting $\hat a_\mu=\hat h_\mu+\sqrt{2}\,\hat p_\mu+\hat r_\mu$ 
where the remaining terms $\hat r_\mu$ containing higher occupation 
numbers do not contribute, we obtain 
\bea
\label{quench}
\langle\hat a_\mu^\dagger(t)\hat a_\nu(t)\rangle
= 
\sum_{\fk{k}} 
e^{i\fk{k}\cdot(\fk{r}_\mu-\fk{r}_\nu)}
\frac{4JUT_{\fk{k}}}{N}
\frac{1-\cos(\omega_{\fk{k}}t)}{\omega^2_{\fk{k}}}
\,,
\eea 
where $N$ denotes the total number of lattice sites.
This prediction could be experimentally verified. 
After preparing bosonic atoms in an optical lattice deep in the 
Mott state $J=0$ and quenching it to a finite $J_{\rm out}=J$, 
time-of-flight images for varying time intervals $t$ after the 
quench yield the Fourier transform of 
$\langle\hat a_\mu^\dagger(t)\hat a_\nu(t)\rangle$ 
and allow the determination of $\omega_{\fk{k}}$ \cite{lattice}. 
In the Mott regime $J_{\rm out}<J_{\rm cr}$, this sweeping method 
allows not only the measurement of the gap value $\omega_{\fk{k}=0}$ 
which is a signature of the insulator phase, but also the $\f{k}$ 
dependence of the energy $\omega_{\fk{k}}$ required for particle-hole 
pair creation.

The expression under the square root in (\ref{omega}) has two zeros 
$J_\pm=U(3\pm\sqrt{8})$ for $k=0$. 
The first one marks the critical point $J_{\rm cr}=J_-$, 
see, e.g., \cite{Navez}. 
For $J<J_{\rm cr}$, i.e., in the Mott phase, 
all modes are stable $\omega_{\fk{k}}\in\mathbb R$.
For $J_{\rm cr}<J<U$ (which is in the superfluid regime),  
all modes with wavenumbers below the value $k_{\rm cr}$  
given by $JT_{k_{\rm cr}}=J_{\rm cr}$ become unstable 
and the mode $k=0$ yields the fastest growth. 
For $J>U$, on the other hand, a finite wave-number $k_*>0$
yields the fastest growth and dominates the evolution of 
$\langle\hat a_\mu^\dagger(t)\hat a_\nu(t)\rangle$,  
see Eq.~(\ref{factorize}) below and \cite{factorize}. 
Finally, for $J>J_+$, long-wavelength modes become stable again 
$\omega_{\fk{k}=0}\in\mathbb R$ and only the modes within a finite 
$k$-interval grow.

In order to search for universal behavior close to the critical point, 
we study a quench not too far into the superfluid regime, i.e., 
$J=J_{\rm cr}(1+\varepsilon)$ with $0<\varepsilon\ll1$.
In this case, the dispersion curve $\omega^2_{\fk{k}}$ dips below zero 
for small $\f{k}$ only, and thus the growth of correlations 
can be determined using the long-wavelength approximation 
\bea
\label{quadratic}
\omega_{\fk{k}}
\approx
i\sqrt{\gamma^2-c^2\f{k}^2}
\,,
\eea 
where $c^2=3J(U-J)/m^*=\ord(Z)$ is a velocity scale.  
Note that the growth rate $\gamma^2\sim J-J_{\rm cr}$ strongly depends 
on the distance to the critical point, whereas $c^2$ is nearly constant. 
For large lattices $N\gg1$, the sum over $\f{k}$ 
is well approximated by an integral.
For large $\gamma t\gg1$, this integral becomes dominated by the 
fastest growing modes and, thus, can be estimated using the saddle-point 
approximation 
\bea
\label{saddle-point}
\langle\hat a_\nu^\dagger(t)\hat a_\mu(t)\rangle 
\approx
{\cal N}(t)
\exp\left\{\gamma\sqrt{t^2-(\f{r}_\mu-\f{r}_\nu)^2/c^2}\right\}
\,,
\eea 
where the time dependence of the normalization factor 
${\cal N}(t)=\ord(1/Z)$ is weak (power-law) compared with the 
exponential growth in (\ref{saddle-point}). 
Focusing on the dominant exponential part in (\ref{saddle-point}),
we find a constant propagation speed $c$ of the correlations 
similar to the Lieb-Robinson bound \cite{Lieb}.  
Furthermore, there is a universal scaling behavior. 
Moving towards or away from the critical point through the 
rescaling $\gamma\to\gamma'=\lambda\gamma$,  
an analogous rescaling of time 
$t\to t'=t/\lambda$ and distance $\f{r}\to\f{r'}=\f{r}/\lambda$ 
leaves the dominant behavior of (\ref{saddle-point}) invariant. 

As another application of (\ref{saddle-point}), let us determine the 
condensate fraction within a compact lattice region $\cal S$ of size 
$|{\cal S}|\gg1$.
By analogy with the continuum case \cite{stringari}, 
the condensate fraction is defined 
via the largest eigenvalue of the two-point correlation function 
$\langle\hat a_\mu^\dagger\hat a_\nu\rangle$. 
Within $\cal S$, the largest eigenvalue corresponds to the homogeneous 
mode which is described by the coarse-grained operator 
$\hat A_{\cal S}=\sum_{\mu\in{\cal S}}\hat a_\mu/\sqrt{|{\cal S}|}$.
For $|{\cal S}|=\ord(Z)$, we obtain a macroscopic occupation 
$N_{\cal S}=\langle\hat A_{\cal S}^\dagger\hat A_{\cal S}\rangle\gg1$
after a finite time $t$.
The precise scaling depends on the size $|{\cal S}|$.
For $|{\cal S}|\ll c^2t^2$, the condensate fraction 
$N_{\cal S}/|{\cal S}|$ is basically independent of $|{\cal S}|$ 
and grows with $\exp\{\gamma t\}$. 
For $|{\cal S}|\gg c^2t^2$, on the other hand, the condensate fraction 
decays $\sim1/|{\cal S}|$.
This suggests that several fragmented condensate patches of size $\ord(ct)$ 
form within the region $\cal S$ which are not yet fully coherent.

These findings motivate the study of the phase correlations. 
To this end, we introduce a coarse-grained phase operator 
$\hat\varphi_{\cal S}$ via 
$\hat A_{\cal S}=\exp\{i\hat\varphi_{\cal S}\}\sqrt{\hat N_{\cal S}}$
with $\hat N_{\cal S}=\hat A_{\cal S}^\dagger\hat A_{\cal S}$. 
For a macroscopic occupation $\langle\hat N_{\cal S}\rangle\gg1$, 
the number fluctuations are negligible so that 
$\hat N_{\cal S}\approx N_{\cal S}$.
Consequently, the phase 
between two non-overlapping regions $\cal S$ and $\cal S'$ of sizes 
$1\ll|{\cal S}|,|{\cal S'}|\ll c^2t^2$ correlates according to 
(for large $\gamma t$)
\bea
\label{phase}
\langle\exp\{i(\hat\varphi_{\cal S}-\hat\varphi_{\cal S'})\}\rangle
\approx 
\exp\left\{-\frac{\gamma\Delta\f{r}^2}{2c^2t}\right\}
\,,
\ea
where $\Delta\f{r}$ is the distance between the two regions.  
The distance over which the phase correlation spreads obeys a 
diffusion-like law $\Delta\f{r}^2\sim c^2t/\gamma$.  
For smaller distances, the relative phases become locked. 

As the final example, the four-point correlation 
\bea
\label{four-point}
\langle\hat a_\mu^\dagger\hat a_\nu^\dagger\hat a_\kappa\hat a_\lambda\rangle
=
\tr_{\mu\nu\kappa\lambda}
\left\{
\hat\rho_{\mu\nu\kappa\lambda}
\hat a_\mu^\dagger\hat a_\nu^\dagger\hat a_\kappa\hat a_\lambda
\right\}
=
\nn
=
\langle\hat a_\mu^\dagger\hat a_\kappa\rangle
\langle\hat a_\nu^\dagger\hat a_\lambda\rangle
+
\langle\hat a_\mu^\dagger\hat a_\lambda\rangle
\langle\hat a_\nu^\dagger\hat a_\kappa\rangle
+
\ord(1/Z^3)
\,,
\ea
is completely determined by (\ref{saddle-point}) 
in leading order $1/Z^2$. 
Despite the similarities, this result is not based on the usual 
Wick expansion (we have a strongly interacting theory) 
but on the hierarchy (\ref{hierarchy}) and the properties of the 
initial Mott state. 
For example, for $\langle\hat a_\kappa\rangle\neq0$, 
we would get an additional contribution of the same order from 
$\hat\rho_{\mu\nu\lambda}^{\rm c}=\ord(1/Z^2)$.  
The above equation allows us to calculate the long-range current 
correlation  
\bea
\label{current}
\langle\hat j_\mu^a(t)\hat j_\nu^b(t)\rangle 
\approx 
\delta_{ab}
\frac{\gamma{\cal N}^2(t)}{2t(m^*c)^2}
\exp\left\{2\gamma t-\frac{\gamma\Delta\f{r}^2_{\mu\nu}}{c^2t}\right\}
\,. 
\ea
%

For the sake of completeness, let us briefly discuss a quench deep 
into the superfluid regime $J>U$, where a finite wavenumber $k_*$ 
yields the fastest growth.
In this case, the correlations behave in a way which is drastically 
different from (\ref{saddle-point}).
The temporal growth and the spatial dependence factorize, 
see also \cite{factorize} 
\bea
\label{factorize}
\langle\hat a_\nu^\dagger(t)\hat a_\mu(t)\rangle 
\approx
{\cal N}_*(t)
\exp\left\{\gamma_* t\right\}
{\cal J}_0(k_*|\f{r}_\mu-\f{r}_\nu|)
\,,
\eea 
with the Bessel function ${\cal J}_0$ and the normalization factor 
${\cal N}_*(t)=\ord(1/Z)$. 
As a result, the correlations grow but do not spread as before in 
(\ref{saddle-point}), i.e., with some velocity. 

In summary, we derived a hierarchy of correlations (\ref{hierarchy}) 
in order to describe the non-equilibrium dynamics of a lattice Bose 
gas (\ref{general}) based on the expansion in inverse powers of the 
large coordination number $1/Z$. 
The lowest order coincides with the Gutzwiller approach, 
cf.~Eq.~(\ref{one-site}), and the higher orders describe the 
correlations, cf.~Eq.~(\ref{two-sites}). 
This method is applied to calculate the creation and amplification 
of quantum correlations in a quenched Mott-superfluid phase transition. 
We find that the off-diagonal long-range order (\ref{saddle-point})
spreads with a constant velocity and obeys universal scaling laws. 
The correlator (\ref{phase}) of the phase associated with local 
condensate patches expands with a diffusion-like law 
(phase locking). 
As an example for higher-order correlations, we calculated the 
four-point function (\ref{four-point}) which yields the current 
correlator (\ref{current}). 

\begin{acknowledgments}
This work was supported by the SFB/TR 12 of the 
German Research Foundation (DFG).
Helpful discussions with I.~Cirac, F.~Queisser, K.~Krutitsky, 
and M.~Pater are gratefully acknowledged.
\end{acknowledgments}



\end{document}